\pdfoutput=1
\newif\ifarxiv
\arxivtrue

\ifarxiv
\documentclass[sigplan,nonacm,screen]{acmart}
\else
\documentclass[sigplan,review]{acmart}
\fi

\ifarxiv
\setcopyright{cc}
\setcctype[4.0]{by}
\else
\renewcommand\footnotetextcopyrightpermission[1]{}
\fi

\usepackage{xcolor,soul}
\usepackage[noline,noend]{algorithm2e}
\usepackage{listings}

\definecolor{GrayCodeBlock}{RGB}{241,241,241}
\definecolor{BlackText}{RGB}{110,107,94}
\definecolor{BlueTypename}{RGB}{17,86,182}
\definecolor{RedTypename}{RGB}{182,86,17}
\definecolor{GreenString}{RGB}{96,172,57}
\definecolor{PurpleKeyword}{RGB}{184,84,212}
\definecolor{GrayComment}{RGB}{170,170,170}
\definecolor{GrayNumber}{RGB}{200,200,200}
\definecolor{GoldDocumentation}{RGB}{180,165,45}
\ifarxiv
\else
\fi

\SetKwProg{CRDT}{crdt}{}{}
\SetKwInput{InitialState}{initialState}
\SetKwProg{Function}{function}{}{}
\SetKwProg{Merge}{merge}{}{}
\SetKwProg{Gossip}{gossip}{}{}
\SetKwProg{Operation}{operation}{}{}
\SetKwProg{Query}{query}{}{}
\SetKwBlock{Loop}{loop}{end}
\SetKw{Break}{break}
\SetKw{Send}{send}

\AtBeginDocument{%
  \providecommand\BibTeX{{%
    \normalfont B\kern-0.5em{\scshape i\kern-0.25em b}\kern-0.8em\TeX}}}

\newif\ifcomments
\commentstrue
\ifcomments
    \providecommand{\shadaj}[1]{{\protect\color{brown}{\bf [shadaj: #1]}}}
    \providecommand{\conor}[1]{{\protect\color{red}{\bf [conor: #1]}}}
    \providecommand{\alvin}[1]{{\protect\color{purple}{\bf [alvin: #1]}}}
    \providecommand{\mae}[1]{{\protect\color{blue}{\bf [mae: #1]}}}
    \providecommand{\joe}[1]{{\protect\color{teal}{\bf [joe: #1]}}}
    \providecommand{\jmh}[1]{{\protect\color{teal}{\bf [joe: #1]}}}
    \providecommand{\david}[1]{{\protect\color{green}{\bf [david: #1]}}}
    \providecommand{\chris}[1]{{\protect\color{violet}{\bf [chris: #1]}}}
    \providecommand{\davidmwei}[1]{{\protect\color{pink}{\bf [david wei: #1]}}}
    \providecommand{\kaushik}[1]{{\protect\color{orange}{\bf [kaushik: #1]}}}
    \providecommand{\justin}[1]{{\protect\color{green}{\bf [justin: #1]}}}
    \providecommand{\mingwei}[1]{{\protect\color{rhodamine}{\bf [mingwei: #1]}}}
    \providecommand{\rithvik}[1]{{\protect\color{red}{\bf [rithvik: #1]}}}
    \providecommand{\nc}[1]{{\protect\color{pink}{\bf [nc: #1]}}}
     \providecommand{\accheng}[1]{{\protect\color{olive}{\bf [accheng: #1]}}}
    \providecommand{\dan}[1]{{\protect\color{purple}{\bf [dan: #1]}}}
\else
    \providecommand{\shadaj}[1]{}
    \providecommand{\conor}[1]{}
    \providecommand{\alvin}[1]{}
    \providecommand{\mae}[1]{}
    \providecommand{\joe}[1]{}
    \providecommand{\jmh}[1]{}
    \providecommand{\david}[1]{}
    \providecommand{\chris}[1]{}
    \providecommand{\davidmwei}[1]{}
    \providecommand{\kaushik}[1]{}
    \providecommand{\justin}[1]{}
    \providecommand{\mingwei}[1]{}
    \providecommand{\rithvik}[1]{}
    \providecommand{\nc}[1]{}
    \providecommand{\accheng}[1]{}
    \providecommand{\dan}[1]{}
\fi

\begin{document}

\title{Optimizing Stateful Dataflow with Local Rewrites}


\author{Shadaj Laddad}
\affiliation{%
 \institution{UC Berkeley}\city{} \country{}}
 \email{shadaj@cs.berkeley.edu}

\author{Conor Power}
\affiliation{%
  \institution{UC Berkeley}\city{} \country{}}
 \email{conorpower@cs.berkeley.edu}

\author{Tyler Hou}
\affiliation{%
 \institution{UC Berkeley}\city{} \country{}}
 \email{tylerhou@berkeley.edu}

\author{Alvin Cheung}
\affiliation{%
 \institution{UC Berkeley}\city{} \country{}}
 \email{akcheung@cs.berkeley.edu}

\author{Joseph M. Hellerstein}
\affiliation{%
 \institution{UC Berkeley}\city{} \country{}}
 \email{hellerstein@cs.berkeley.edu}

\begin{abstract}
Optimizing a stateful dataflow language is a challenging task. There are strict correctness constraints for preserving properties expected by downstream consumers, a large space of possible optimizations, and complex analyses that must reason about the behavior of the program over time. Classic compiler techniques with specialized optimization passes yield unpredictable performance and have complex correctness proofs. But with e-graphs, we can dramatically simplify the process of building a correct optimizer while yielding more consistent results! In this short paper, we discuss our early work using e-graphs to develop an optimizer for a the Hydroflow dataflow language. Our prototype demonstrates that composing simple, easy-to-prove rewrite rules is sufficient to match techniques in hand-optimized systems.
\end{abstract}


\maketitle

\section{Introduction}
As applications scale to handle geodistributed users who interact in real-time, streaming dataflow systems have gained popularity as a way to enable low-latency computations on live data. Existing dataflow systems focus primarily on execution performance, utilizing incremental computation~\cite{cql, naiad} and operator fusion~\cite{weld}. But recent directions such as the Hydro project~\cite{hydro} focus on language designs that make it easier for developers to reason about correctness and compilers to \emph{automatically} discover and apply optimizations.

In this extended abstract, we focus on our efforts in building an optimizer for Hydroflow~\cite{hydroflow,hydroflow-thesis}, a low-level dataflow language embedded in Rust. Hydroflow is intended as an ``LLVM IR for distributed programs,'' designed to provide a simple and clear execution model that can be leveraged as a target of higher-level languages. For example, we are actively building a execution engine for Dedalus~\cite{dedalus}, a variant of Datalog, by compiling it down to Hydroflow.

Being at the lowest level of the Hydro stack, individual Hydroflow programs are actually \emph{not distributed}---they describe a streaming computation executed on a single thread. Instead, Hydroflow focuses on making effects of \emph{time}, such as non-deterministic batching, explicit in the program. Hydroflow programs can then be safely composed with network connections at their boundaries to form a distributed system.

Writing an optimizer for such a programming language is a daunting task. We need to apply program-wide transformations in the style of a query optimizer~\cite{cascades}, but using heuristics to order optimization passes can lead to unpredictable performance. Many Hydroflow transformations result in graphs with equal or higher intermediate cost, but can enable later optimizations that dramatically reduce the final cost. Because Hydroflow is a compiler target, ordered passes are especially problematic because they would place a burden on upstream compilers to emit "optimizer-friendly" Hydroflow. 

But e-graphs~\cite{e-graph,e-graph-2} give us a glimmer of hope! Instead of greedily making optimization decisions, we can compose local rewrite rules and efficiently explore the full space of transformations. Using e-graphs to drive our optimizing compiler enables three key opportunities:
\begin{enumerate}
\item We can define primitive rewrites that map to core dataflow properties (distributive, deterministic, etc.) instead of brittle special-cases.
\item Our correctness proofs are much simpler, because we can independently prove low-level rules.
\item We can implement optimizations that involve \emph{inductive} proofs over \emph{time}, by using equivalence predicates that search the e-graph for cycles.
\end{enumerate}


We present our early work using e-graphs in Hydroflow to optimize stateful dataflow. Our rewrite rules make special use of the e-graph model, with some rules searching for \emph{equivalence cycles} that can be optimized into incremental computation. By applying local rewrites, Hydroflow can expose high-level stateful operators while ensuring that they are optimized away ahead-of-time. In particular, our rewrite rules are sufficient for the optimizer to automatically discover classic patterns such as streaming joins.

Our rewrite rules are easy to verify, and can be applied in a general-purpose manner to any dataflow. Although we have seen success in our early prototypes, e-graphs are not (yet) a perfect solution for dataflow optimization. We discuss diamonds, a class of dataflow structures that are difficult to optimize with e-graphs, and explore directions for future work to address these limitations.

\section{Motivating Example}
Before we dive into the optimizer, let us explore how developers can build streaming dataflow services in Hydroflow. Consider a simple chat application, where users can join a channel and receive all messages sent (including those before they joined). To keep things simple, we'll consider the case where there is only a single channel.

Hydroflow programs are written sets of declarative statements that connect \textbf{pipelines} to each other through \textbf{operators}, which define logic such as \texttt{map}, \texttt{filter}, or \texttt{join}. Operators can take multiple inputs (senders explicitly index into these), and local pipelines can be created by chaining together several operators. In addition, Hydroflow supports dataflow cycles, which if present are run to fixpoint.

Our application has two streaming inputs, one for users requesting to join the channel (\texttt{add\_member}), and one for messages being sent by users (\texttt{messages}). We can send messages to users by sending $(\mathit{user}, \mathit{msg})$ pairs to a downstream \texttt{notify} pipeline. An initial attempt to implement this in Hydroflow may look like the following.

\begin{lstlisting}
add_member -> [0] broadcast
messages -> [1] broadcast
broadcast = cross() -> notify
notify = ...
\end{lstlisting}

In this case, we can broadcast messages by taking the cross product (with \texttt{cross}) of the users added to the channel and messages sent. But this program will actually behave incorrectly! To understand why, we need to dive into Hydroflow's execution model.

Each Hydroflow program (``spinner'') executes as an event loop. At the beginning of each iteration, called a ``tick,'' Hydroflow collects any available network packets for each input channel into a batch.
The dataflow is then executed on these batches, and once fixpoint is reached the values accumulated at each output are flushed to the network. Critically, all dataflow operators are \emph{stateless} by default, so all state is cleared at the end of a tick.

In our example, this means that our program will only broadcast messages to users that joined \emph{in the same tick}. This ``catch'' is by design---Hydroflow guides users to be mindful of the effects of network latency and batching on their programs. Indeed, there is nothing in our code that corresponds to showing previous messages to newly joined users.

Let us fix this. Hydroflow has a \emph{stateful} operator, \texttt{persist}, which consumes elements from some upstream source and emits the \emph{entire history} of values it has received up to and including the current tick. With this operator, it is easy to get a more sensible program:

\begin{lstlisting}
add_member -> persist() -> [0] broadcast
messages -> persist() -> [1] broadcast
broadcast = cross() -> notify
\end{lstlisting}

There is still one last issue. Because \texttt{persist} replays the entire history of messages in every tick, clients will be sent repeated notifications for the same message. We can fix this by using the inverse of \texttt{persist}, the \texttt{delta} operation. This dataflow element consumes values from an upstream source, but only emits the \emph{new} values in this tick. So our final, complete program looks like the following:

\begin{lstlisting}
add_member -> persist() -> [0] broadcast
messages -> persist() -> [1] broadcast
broadcast = cross() -> delta() -> notify
\end{lstlisting}

That's all! We now have a precise implementation of our specified program semantics. But this is not particularly efficient. In a naive execution of this dataflow, we will take the cross product with all messages in the history of the channel, only to later perform a delta that retains only the new messages and replays for newly joined members. Our goal is to preserve this clear model for computation while optimizing away the inefficiencies of naive state accumulation.

\section{Optimizing Stateful Dataflow}
\label{sec:optimization}
To tackle the issue of inefficient stateful operators in dataflows, we turn to e-graphs. Our goal is to identify rewrite rules that optimize subflows while preserving which values are emitted and any ordering guarantees. An important principle in our usage of e-graphs is boiling down optimizations to first principles. Rather than baking in specific rules for operations like cross-products, we instead want to identify more general rules that can be composed during e-graph expansion.

First, we need to define an encoding of Hydroflow graphs as expressions that we can define rewrite rules over. For our prototype, we use a tree encoding, where dataflow operators are defined as functions with inputs passed as parameters. In our prototype, we elide any non-dataflow inputs (such as user-defined functions) because our optimizations do not currently make use of them. With our encoding, the motivating example can be expressed as an expression:

\begin{lstlisting}
(delta (cross
  (persist add_member)
  (persist messages)))
\end{lstlisting}

Using a tree encoding has some limitations, such as being unable to express dataflows where a shared computation has several consumers. We discuss our current solutions for these challenges and propose opportunities for future research in Section~\ref{section:diamonds}. In our discussion of using e-graphs to discover dataflow optimizations, this representation is sufficient.

\subsection{Rewriting Persist}
Let's start with some of the simpler rules. In the previous section, we introduced the \texttt{persist} and \texttt{delta} operators for reasoning about accumulated state. These are inverses, so we can define a rewrite for \texttt{persist} followed by a \texttt{delta}. In the syntax of egg, we can specify the rewrite:

\begin{center}
\textbf{\texttt{(delta (persist ?a)) <=> ?a}}
\end{center}

Next, we develop rewrite rules for reasoning about the behavior of \texttt{persist}. We discussed earlier that \texttt{persist} replays the messages it received in previous ticks, and also emits the values received from upstream in the current loop. A natural rewrite rule, then, is to make these semantics explicit so that our optimizer can reason about these two sources of values.

We can introduce a new operator \texttt{old}, which behaves the same as \texttt{persist} \emph{except} it does not emit the new values received from upstream. Then we can use the \texttt{chain} operator, which combines messages received from two upstream channels by emitting all values from the first before all values from the second, to rewrite a \texttt{persist}:

\begin{center}
\textbf{\texttt{(persist ?a) <=> (chain (old ?a) ?a)}}
\end{center}

With the rules so far, we can rewrite our working example to replace the \texttt{persist} operators:

\begin{lstlisting}
(delta (cross
  (chain
    (old add_member)
    add_member)
  (chain
    (old messages)
    messages)))
\end{lstlisting}

\subsection{Distributing Cross Products}
A natural next step for our rewrite rules is to reason about cross products over chained input channels. The cross product operator is distributive over chains (it makes no guarantees about the order of output tuples), so we can define a rewrite rule for this. Because \texttt{cross} is not commutative (the order of elements \emph{in} each tuple matters), we must also define a rule for when the chain is in the second input. In summary, we add the following rules:

\begin{center}
\textbf{\texttt{(cross (chain ?a ?b) ?c) <=> (chain (cross ?a ?c) (cross ?b ?c))}}

\textbf{\texttt{(cross a? (chain ?b ?c)) <=> (chain (cross ?a ?b) (cross ?a ?c))}}
\end{center}

Applying both of these rules to our working example, we can shift both chain operators to the other side of the cross product, which reveals how new and old data individually contribute to the final result:

\begin{lstlisting}
(delta (chain
  (chain
    (cross
      (old add_member) (old messages))
    (cross (old add_member) messages)
  )
  (chain
    (cross add_member (old messages))
    (cross add_member messages)
  )))
\end{lstlisting}

Next, we have another rewrite rule corresponding to a fundamental property of dataflow operators: \emph{associativity}. Because \emph{chain} is associative, we can shift around the grouping with a rewrite rule:

\begin{center}
\textbf{\texttt{(chain (chain ?a ?b) ?c) <=>}}

\textbf{\texttt{(chain ?a (chain ?b ?c))}}
\end{center}

This rewrite rule allows us to isolate the cross product dealing with only old values, which we will need later on to make this computation incremental:

\begin{lstlisting}
(delta (chain
  (cross
    (old add_member) (old messages))
  (chain
    (cross (old add_member) messages)
    (chain
      (cross add_member (old messages))
      (cross add_member messages)
    )
  )
))
\end{lstlisting}

\subsection{Modeling Determinism}
Our next insight is that we have not yet used the property of \emph{determinism} in a rewrite. We know that \texttt{cross} (along with most other dataflow operators) is deterministic--it produces the same tuples (still with no ordering guarantee) over ticks as long as the input streams produce the same values.

Let us codify this by introducing a new dataflow operator \texttt{prev}. This operator simply emits the values it received in the \emph{previous} tick. First, we define a rewrite relating \texttt{old} and \texttt{persist} with \texttt{prev}. Then, we can define a rewrite rule for \texttt{cross} that uses the fact that it is deterministic to shift a computation to a previous tick:

\begin{center}
\textbf{\texttt{(old ?a) <=> (prev (persist ?a))}}

\textbf{\texttt{(cross (prev ?a) (prev ?b)) <=>}}

\textbf{\texttt{(prev (cross ?a ?b))}}
\end{center}

Again, these rewrite rules describe the core properties of our operators rather than a specific optimization case. Let's take a look at a rewrite of our program with these rules applied:

\begin{lstlisting}
(delta (chain
  (prev (cross
    (persist add_member) (persist messages)
  ))
  (chain
    (cross (old add_member) messages)
    (chain
      (cross add_member (old messages))
      (cross add_member messages)
    )
  )
))
\end{lstlisting}

\subsection{Putting it Together: Incrementalization}
Finally, we we can optimize our dataflow into an incremental computation. We notice that the e-node for the subexpression inside the \texttt{delta} has remained within the same e-class as the original subexpression (\texttt{(cross (persist add\_member) (persist messages))}, at the beginning of Section~\ref{sec:optimization}). This original subexpression appears \emph{within} our \texttt{chain}, which means that we could instead just add to the existing result from the previous tick! This is an exciting result; we have identified an incremental way to compute the cross product by composing primitive rewrites rather than writing specialized rules.

Indeed, we only have one rewrite that deals with incremental computation. We are looking for a cycle through a \texttt{prev} node, so we can attach a predicate that checks for equivalence between the root and the child inside \texttt{prev}:

\begin{center}
\textbf{\texttt{(chain (prev ?a) ?b) => (persist ?b)}}

\textbf{\texttt{if eclass((chain (prev ?a) ?b)) = eclass(?a)}}
\end{center}

The proof of correctness for this rewrite relies on induction over ticks. In the base case, \texttt{(prev ?a)} is an empty stream, so \texttt{(chain (prev ?a) ?b)} = \texttt{?b} = \texttt{(persist ?b)} because there are no previous persisted values. In the inductive step, we know that \texttt{(prev (chain (prev ?a) ?b))}. Then, our equivalence constraint says that \texttt{(prev (persist ?b))} = \texttt{(prev (chain (prev ?a) ?b))} = \texttt{(prev ?a)}. We can wrap these expressions to get \texttt{(chain (prev ?a) ?b)} = \texttt{(chain (prev (persist ?b)) ?b)}. The latter is the definition of \texttt{(persist ?b)} so our rule is correct. Applying this to our working example, we get:

\begin{lstlisting}
(delta (persist (chain
  (cross (old add_member) messages)
  (chain
    (cross add_member (old messages))
    (cross add_member messages)
  )
)))
\end{lstlisting}

Finally, we can apply the first rewrite rule we defined to cancel out the \texttt{delta} and \texttt{persist} and obtain an efficient, incremental dataflow:

\begin{lstlisting}
(chain
  (cross (old add_member) messages)
  (chain
    (cross add_member (old messages))
    (cross add_member messages)
  )
)
\end{lstlisting}

So far, we have discussed only the rewrite rules, but have not specified how we pick a single rewritten program from the expanded e-graph. To do this, we can specify a simple cost model that computes the number of nodes with a higher weight for \texttt{delta} nodes because these indicate duplicated work. With this cost model, we can apply the same set of rewrite rules to a three-way cross product (between \texttt{add\_member}, \texttt{messages}, and \texttt{platforms}), and discover an appropriate incremental algorithm:

\begin{lstlisting}
(chain
  (cross
    add_member
    (cross
      (old messages) (old platforms)))
  (cross
    (persist add_member)
    (chain
      (cross messages (old platforms)) 
      (cross
        (persist messages) platforms))))
\end{lstlisting}

What is exciting is that our rules around manipulating \texttt{delta}/\texttt{persist}/\texttt{old} operators are general, with no rules specific to the \texttt{cross} case. If we define similar rules for distribution and determinism over the \texttt{join} operation, we can derive the incremental semi-naive datalog evaluation from scratch! By using e-graphs to explore the search space of composed rewrites, we are able to easily support a large swath of programs with minimal effort needed to define rewrites and verify their correctness.




\section{Diamonds are Hard to Crack}
\label{section:diamonds}
There is one limitation of our approach using e-graphs that is hard to ignore, yet leaves many exciting opportunities for future work in the wider e-graphs space. In Hydroflow, the data flowing out of a node can be used by several downstream paths through the \texttt{tee} operator, which at runtime sends copies of each incoming value to each consuming operator. For example, we can use \texttt{tee} to compute users who should meet at the next Bay Area e-graph meetup:

\begin{lstlisting}
members = add_member -> persist() -> tee()
meetup = cross()
members -> map(with_school)
  -> filter(berkeley) -> [0] meetup
members -> map(with_school)
  -> filter(stanford) -> [1] meetup
\end{lstlisting}

In the cost model for an optimizer, it is critical to take into account that the computation before the \text{tee} is only performed once each tick, regardless of the number of consumers. But with our encoding of dataflow as expressions, this is currently not possible, because we can only extract computation trees rather than general DAGs.

In particular, the dataflow structure that breaks our encoding is a \textbf{diamond}, a dataflow where a common computation is transformed in different ways that are eventually merged together (by interleaving their elements, joining on a key, etc). This is similar to a common table expression (CTE) in database lingo or a let-binding in functional languages, where a single result is produced.

In our current prototype, we simply flatten all diamonds by duplicating their shared subexpressions, and re-form diamonds after optimization by searching for identical expressions in the output. But in an ideal system, diamonds would be handled just like any other constructs in the optimizer. There are three key challenges in optimizing diamonds:

\begin{enumerate}
\item When computing the cost function for a node, a common subexpression's cost should be counted only once even if is referenced multiple times.
\item We may want to shift logic from the common subexpression into its downstream consumers (inlining), to enable further optimizations.
\item The reverse of (2), after performing rewrites we may want to extract shared logic into a common subexpression to avoid duplicate computation.
\end{enumerate}

\subsection{Forming Diamonds with Zippers}
In our early prototypes, we designed an explicit operator that captures the structure of a diamond. The \texttt{diamond} operator takes four parameters: the shared computation, two ``edges'' that describe the transformations being applied to data from the common source, and a merge node that defines how to combine the results from the two edges. This representation immediately solves challenge (1), since we precisely capture which computation is shared between multiple paths.
For example, we can encode the earlier example with \texttt{diamond}:

\begin{lstlisting}
(diamond
  (persist add_member)
  (zipper
    in
    (map with_school 
      (filter berkeley out)))
  (zipper
    (filter stanford 
      (map with_school in))
    out)
  (cross first second))
\end{lstlisting}

A key trick in this formulation is representing the ``edges'' of the diamond using a zipper~\cite{zipper} data structure. The nesting of operators is reversed between the halves of the zipper. In the first half, operator nodes have their inputs as children, but in the second half they have consumers as children. We use two special variables, \texttt{first} and \texttt{second}, to reference the values flowing out of both edges of the diamond. 

What is powerful about zippers is that they make it possible to isolate either the first \emph{or} last operator in a sequence by shifting the ``cursor,'' the point where the two halves meet. In standard zipper implementations, this is implemented by popping an element from one half and pushing it to the other. For our encoding, we similarly remove the outermost operator from one half and wrap the other half with it. In our example, we can shift the cursors in both zippers to isolate one operator in each half:

\begin{lstlisting}
(diamond
  (persist add_member)
  (zipper
    (map with_school in)
    (filter berkeley out))
  (zipper
    (map with_school in)
    (filter stanford out))
  (cross first second))
\end{lstlisting}

After isolating the last operator of the edge in the second half of a zipper, we can apply another rewrite to inline the operator in the output, solving challenge (2). Thanks to the symmetry of the zipper, we can \emph{also} solve challenge (3). If a single operator is isolated in the \emph{first} half of each zipper, and is the same for both edges, it can be shared. Applying both rewrites, we can transform our example to:

\begin{lstlisting}
(diamond
  (map with_school (persist add_member))
  (zipper in out)
  (zipper in (filter stanford out))
  (cross (filter berkeley first) second))
\end{lstlisting}

This encoding comes with a catch: there are dataflow graphs that \emph{cannot} be encoded in terms of this diamond operator. Because our zippers only represent flat sequences, we cannot have any multi-input operators along an edge, unless those operators are part of a sub-diamond. 
In addition, manipulating zippers is very expensive, as we generate new e-classes for both halves whenever we perform a cursor shift, causing the e-graph to expand quite quickly. But rewrite rules \emph{consuming} zippers only care about the \emph{isolated} first or last operator, so other intermediate states only exist for the shifting rule. In future work, we hope to explore ways to more efficiently represent zipper structures in an e-graph to take advantage of this domain-specific knowledge rather than naively using standard rewrite rules.






\section{Conclusion}
Developing optimizers is a challenging task, and building one for a low-level dataflow language is that much more daunting. The space of possible optimized programs is massive, and developing specialized rules can lead to brittle behavior. But with e-graphs, we can boil down dataflow optimization into a set of core rules that map to fundamental properties of operators such as associativity and determinism. By leveraging the composition of these rules, we can automatically discover optimizations such as incremental joins without specialized rules and cumbersome proof effort. E-graphs are not a perfect solution for all cases, with diamonds particularly hard to optimize, but there are promising directions that allow us to preserve the simplicity of local rewrites while supporting more programs.

\begin{acks}
We thank our anonymous reviewers for their insightful feedback on this paper. This work is supported in part by National Science Foundation CISE Expeditions Award CCF-1730628, IIS-1955488, IIS-2027575, DOE award DE-SC0016260, ARO award W911NF2110339, and ONR award N00014-21-1-2724, and by gifts from Amazon Web Services, Ant Group, Ericsson, Futurewei, Google, Intel, Meta, Microsoft, Scotiabank, and VMware. Shadaj Laddad is supported in part by the NSF Graduate Research Fellowship Program under Grant No. DGE 2146752. Any opinions, findings, and conclusions or recommendations expressed in this material are those of the authors and do not necessarily reflect the views of the National Science Foundation.
\end{acks}

\bibliographystyle{ACM-Reference-Format}
\bibliography{bibliography}

\end{document}